\documentclass[a4paper]{article}

\usepackage{INTERSPEECH2021}
\usepackage{multirow}
\usepackage{cite}

\title{Channel-wise Gated Res2Net: Towards Robust Detection of \\ Synthetic Speech Attacks}
\name{Xu Li$^1$, Xixin Wu$^2$, Hui Lu$^1$, Xunying Liu$^1$, Helen Meng$^1$}
\address{
  $^1$Department of Systems Engineering and Engineering Management, \\ The Chinese University of Hong Kong \\
  $^2$Department of Engineering, University of Cambridge}
  
\email{\{xuli, luhui, xyliu, hmmeng\}@se.cuhk.edu.hk, xw369@cam.ac.uk}

\begin{document}

\maketitle

\begin{abstract}
Existing approaches for anti-spoofing in automatic speaker verification (ASV) still lack generalizability to unseen attacks. The Res2Net approach designs a residual-like connection between feature groups within one block, which increases the possible receptive fields and improves the system's detection generalizability. However, such a residual-like connection is performed by a direct addition between feature groups without channel-wise priority. We argue that the information across channels may not contribute to spoofing cues equally, and the less relevant channels are expected to be suppressed before adding onto the next feature group, so that the system can generalize better to unseen attacks. This argument motivates the current work that presents a novel, channel-wise gated Res2Net (CG-Res2Net), which modifies Res2Net to enable a channel-wise gating mechanism in the connection between feature groups. This gating mechanism dynamically selects channel-wise features based on the input, to suppress the less relevant channels and enhance the detection generalizability. Three gating mechanisms with different structures are proposed and integrated into Res2Net. 
Experimental results conducted on ASVspoof 2019 logical access (LA) demonstrate that the proposed CG-Res2Net significantly outperforms Res2Net on both the overall LA evaluation set and individual difficult unseen attacks, which also outperforms other state-of-the-art single systems, depicting the effectiveness of our method.

\end{abstract}


\section{Introduction}
Spoofing attacks on automatic speaker verification (ASV) have attracted ever-increasing security concerns in recent years, as they pose serious threats to essential applications of ASV, such as e-banking authentication, device activation, etc. These attacks can be categorized into human impersonation \cite{vestman2020voice,kinnunen2019can}, audio replay \cite{wu2015spoofing,chettri2018study}, synthetic speech \cite{shchemelinin2013examining,kinnunen2012vulnerability} and the recently emerged adversarial attacks \cite{das2020attacker,wu2020defense,li2020adversarial,wu2021adversarial,peng2021pairing,wu2021improving}.

In the midst of the arms race between attack and defense for ASV, the speech community has held several ASVspoof Challenges \cite{wu2015asvspoof,kinnunen2017asvspoof,todisco2019asvspoof} to develop countermeasures mainly against audio replay, text-to-speech (TTS) and voice conversion (VC) attacks. ASVspoof 2019 \cite{todisco2019asvspoof} is the latest one that contains two sub-challenges: physical access (PA) and logical access (LA). PA considers spoofing attacks from replay while LA refers to attacks generated with TTS and VC techniques. 

A model's generalizability to unseen spoofing attacks is challenging but essential for developing reliable countermeasures \cite{li2020investigating,nautsch2021asvspoof}.
% which is also one of the main evaluation perspectives in ASVspoof 2019. The unseen spoofing attacks exist in the evaluation partition where replay configurations for PA and spoofing algorithms for LA differ from those in the training and development partitions.
% Detection of unseen PA attacks can be largely solved by design of powerful system architectures \cite{cheng2019replay,li2021replay,lavrentyeva2019stc}, generalized acoustic features \cite{cheng2019replay,cai2019dku}, etc.
To tackle this issue, previous efforts dedicated to the design of powerful system architectures \cite{cheng2019replay,li2021replay,lavrentyeva2019stc} and generalized acoustic features \cite{cheng2019replay,cai2019dku} have shown great enhancement of generalization to unseen PA attacks.
However, the unseen nature of LA attacks has larger variations due to numerous available speech synthesis algorithms,  and some of them are difficult to be detected, e.g. the A17 attack in the LA evaluation set \cite{todisco2019asvspoof}.
State-of-the-art (SOTA) countermeasures may easily overfit to the training and development sets, and lack good generalizability to unseen LA attacks \cite{nautsch2021asvspoof}. Hence, this work focuses on enhancing generalized detection of LA attacks.

Much promising effort has been dedicated to designing countermeasures against LA attacks \cite{wu2012detecting,alzantot2019deep,lai2019assert,lavrentyeva2019stc,zhang2021one}. Das et al. \cite{das2021data} augments the training data based on signal companding methods to enhance generalization. The RawNet2 architecture \cite{tak2021end} is applied to detect synthetic speech directly upon the raw speech waveform. Our earlier work \cite{li2021replay} leverages the Res2Net architecture to improve the model's generalizability and demonstrates its superior detection accuracy on unseen LA attacks.

\begin{figure*}[th]
    \centering
    \includegraphics[width=\textwidth]{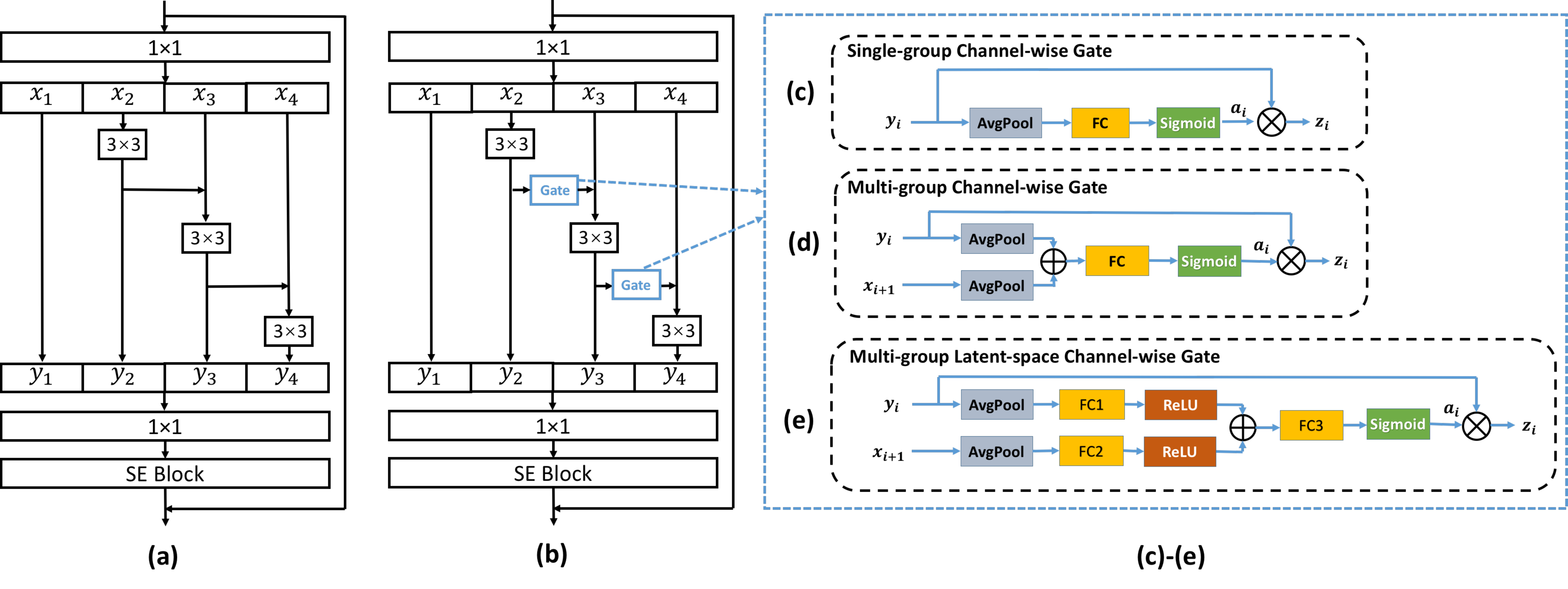}
    \caption{Illustration of different structures in the modules: (a) Res2Net block; (b) CG-Res2Net block; (c) Single-group Channel-wise Gate; (d) Multi-group Channel-wise Gate; (e) Multi-group Latent-space Channel-wise Gate. (SE Block: the squeeze-and-excitation block \cite{hu2018squeeze}; AvgPool: average pooling; FC: fully-connected layer; ReLU and Sigmoid are the two activation functions.)}
    \label{fig:network_structure}
    % \vspace{-0.3cm}
\end{figure*}

According to \cite{li2021replay}, Res2Net designs a residual-like connection between different feature groups within one block, which increases the possible receptive fields and helps the system generalize better to unseen attacks. However, such a residual-like connection is performed by a direct addition between feature groups without channel-wise priority. We argue that the information across channels within a group may not contribute to spoofing cues equally, and the less relevant channels are expected to be suppressed before adding to another feature group, so that the system can generalize better to unseen attacks.

From the above motivation, this work extends \cite{li2021replay} and proposes a novel network architecture, i.e. channel-wise gated Res2Net (CG-Res2Net). CG-Res2Net modifies the Res2Net block and enables a channel-wise gating mechanism in the residual-like connection between feature groups. This gating mechanism selects the more relevant channels while suppresses the less relevant ones to dynamically adapt to unseen attacks, hence enhances the model's detection generalization.
% As an extension of \cite{li2021replay}, this work proposes a novel network architecture, i.e. channel-wise gated Res2Net (CG-Res2Net), which integrates a channel-wise gating mechanism into the Res2Net block. Res2Net designs a residual-like connection between different feature groups within one block, which increases the possible receptive fields and helps the system generalize better to unseen attacks \cite{li2021replay}. However, such a residual-like connection is performed by a direct addition between feature groups without channel-wise priority. We argue that information of each channel within one feature group may be not equally contributed to spoofing cues, even some channels may provide no spoofing cues. Such channel information that is unrelated to spoofing cues, is expected to be depressed, so that the system can generalize better to unseen attacks. Motivated by this, this work proposes CG-Res2Net, which integrates a channel-wise gating mechanism into the Res2Net block, which provides an ability of channel-wise feature selection to depress unexpected channel information. 
% While Res2Net improves generalization via a residual-like connection between different feature groups, the connection is a direct addition without channel-wise priority. We argue that information of each channel within a group may not be equally important to capture spoofing cues, such that a channel-wise gating mechanism is essential to benefit capturing spoofing cues and aid generalization. 
Specifically, we propose and compare three possible gating mechanisms: single-group channel-wise gate (SCG), multi-group channel-wise gate (MCG) and multi-group latent-space channel-wise gate (MLCG). SCG automatically learns a channel-wise gate by a fully-connected layer, based on the current feature group. MCG differs from SCG by additionally considering the next feature group information as reference to compute the gate. Finally, MLCG modifies MCG to firstly project the features of two groups into separate latent spaces, then compute the gate based on the two latent spaces. Three gating mechanisms are integrated with Res2Net to form SCG-Res2Net, MCG-Res2Net and MLCG-Res2Net, respectively. The proposed systems are evaluated on the ASVspoof 2019 LA partition in terms of the performance on the overall LA evaluation set as well as individual difficult unseen attacks. Experimental results demonstrate the effectiveness of the proposed gating mechanisms. All three proposed CG-Res2Net models outperform other single, SOTA systems on ASVspoof 2019 LA evaluation, depicting the promising performance of the CG-Res2Net models.

The contributions of this work include: 1) Proposing a novel CG-Res2Net architecture which can integrate one of three different channel-wise gating mechanisms into the Res2Net block; 2) Demonstrating that three proposed CG-Res2Net models outperform Res2Net on the overall LA evaluation set as well as individual difficult unseen attacks; 3) The proposed CG-Res2Net models outperform other single, SOTA systems on the ASVspoof 2019 LA evaluation set.

The rest of this paper is organized as follows: Section~\ref{sec:channel-wise-att-res2net} illustrates the proposed CG-Res2Net architecture and three gating mechanisms. Experimental setup and results are demonstrated in Section~\ref{sec:expt-setup} and Section~\ref{sec:expt-rst}, respectively. Finally, Section~\ref{sec:conclusion} concludes this work.

\section{Approach}
\label{sec:channel-wise-att-res2net}

\subsection{Channel-wise gated Res2Net}
This section introduces the network structure of proposed CG-Res2Net. CG-Res2Net modifies the Res2Net block to enable a channel-wise gating mechanism in the residual-like connection between feature groups. The comparison between the structures of Res2Net and CG-Res2Net blocks is illustrated in Fig.~\ref{fig:network_structure} (a) and (b). After a $1 \times 1$ convolution, both models evenly split the input feature map $X$ by the channel dimension into $s$ subsets, denoted by $x_i$, where $i \in \{1,2,...,s\}$. We assume that $X \in \mathbb{R}^{sC \times D \times T}$ and each $x_i \in \mathbb{R}^{C \times D \times T}$, where $C$, $D$ and $T$ denote the dimensions of channel, spectrum and time, respectively.

Res2Net enables a direct addition between feature groups before a $3 \times 3$ convolution. Each $y_i$ is derived as Eq.~\ref{eq:res2net-block}:
\begin{align}
\label{eq:res2net-block}
    y_i = \begin{cases}
     x_i, & i=1 \\
     K_{i}(x_i), & i=2 \\
     K_{i}(x_i+y_{i-1}), & 2 < i \leq s
    \end{cases}
\end{align}
where each $K_i()$ denotes a convolutional function with a parameter size of $3 \times 3$. CG-Res2Net adopts a gating mechanism in the residual-like connection. Each $y_i$ is derived as follows:
\begin{align}
    y_i &= \begin{cases}
     x_i, & i=1 \\
     K_{i}(x_i), & i=2 \\
     K_{i}(x_i+z_{i-1}), & 2 < i \leq s
    \end{cases} \label{eq:CG-Res2Net-block-1} \\
    z_{i-1} &= y_{i-1} \otimes a_{i-1} \label{eq:CG-Res2Net-block-2}
\end{align}
where $z_{i}$ scales $y_i$ by a channel-wise gate $a_i \in \mathbb{R}^{C}$, and $\otimes$ denotes a channel-wise multiplication operation. We expect that such an gating mechanism gives priority to channels that contain most spoofing cues and suppresses the less relevant channels, then enhances the model's generalizability to unseen attacks.

This work proposes three novel channel-wise gating modules to be integrated with the Res2Net block, as shown in Fig.~\ref{fig:network_structure} (c)-(e). The detailed functionality of each module is demonstrated in Section~\ref{subsec:channel-wise-att}. Our codes have been made open-source\footnote{https://github.com/lixucuhk/Channel-wise-Gated-Res2Net}.

\subsection{Channel-wise gating mechanism}
\label{subsec:channel-wise-att}

\subsubsection{Single-group channel-wise gate}
As shown in Fig.~\ref{fig:network_structure} (c), the single-group channel-wise gate (SCG) automatically learns a gate $a_i$ given the current feature group $y_i$. The mapping is achieved by a fully-connected layer. $y_i$ is firstly squeezed to the channel dimension by averaging over the spectrum and time dimensions (Eq.~\ref{eq:sca-att-avgpool}), and then transformed by a fully-connected layer $W_{fc} \in \mathbb{R}^{C \times C}$ with a sigmoid activation function $\sigma$ to derive the gate $a_i$ (Eq.~\ref{eq:sca-att-final}).
\begin{align}
    F_{ap}(y_i) &= \frac{1}{D \times T}\sum_{d=1}^{D}\sum_{t=1}^{T}y_i(:,d,t) \label{eq:sca-att-avgpool} \\
    a_i &= \sigma [W_{fc}^{T}F_{ap}(y_i)]
    \label{eq:sca-att-final}
\end{align}

\subsubsection{Multi-group channel-wise gate}
Since the residual-like connection is operated between $y_i$ and $x_{i+1}$, it may be helpful to consider $x_{i+1}$ as reference when applying the gating mechanism. Thus we propose the multi-group channel-wise gate (MCG) where the channel-wise gate is derived from both information of $y_i$ and $x_{i+1}$, as shown in Fig.~\ref{fig:network_structure} (d). This is formulated as Eq.~\ref{eq:mca-att-final}:
\begin{align}
    a_i = \sigma \{W_{fc}^{T}[F_{ap}(y_i) \oplus F_{ap}(x_{i+1})]\}
    \label{eq:mca-att-final}
\end{align}
where $\oplus$ is a concatenation function. $y_i$ and $x_{i+1}$ are squeezed to the channel dimension by $F_{ap}$, then concatenated together and transformed by a linear matrix $W_{fc} \in \mathbb{R}^{2C \times C}$ with sigmoid activation to derive $a_i$.

\subsubsection{Multi-group latent-space channel-wise gate}
With the consideration that $x_{i+1}$ provides information as reference while $y_i$ contains information to be re-scaled, the functionalities of them are not symmetric and it may be better to process them independently before concatenation. Thus we propose the multi-group latent-space channel-wise gate (MLCG) that separately project $y_i$ and $x_{i+1}$ into each own latent space before concatenation, as shown in Fig.~\ref{fig:network_structure} (e). Moreover, to limit model complexity and aid generalization, the latent space could have a reduced dimension with reduction ratio $r$. Specifically, $y_i$ and $x_{i+1}$ are squeezed by $F_{ap}$, then transformed by $W_{fc1} \in \mathbb{R}^{C \times \frac{C}{r}}$ and $W_{fc2} \in \mathbb{R}^{C \times \frac{C}{r}}$ with ReLU activation $\delta$, respectively, as shown in Eq.~\ref{eq:mlca-att-l1-trans} and \ref{eq:mlca-att-l2-trans}. The squeezed channel information is concatenated together, to be transformed by $W_{fc3} \in \mathbb{R}^{\frac{2C}{r} \times C}$ with sigmoid activation to derive $a_i$, as shown in Eq.~\ref{eq:mlca-att-final}.

\begin{align}
    &L_1(y_i) = \delta(W_{fc1}^{T}F_{ap}(y_i)) \label{eq:mlca-att-l1-trans} \\
    &L_2(x_{i+1}) = \delta(W_{fc2}^{T}F_{ap}(x_{i+1})) \label{eq:mlca-att-l2-trans} \\
    &a_i = \sigma \{W_{fc3}^{T}[L_1(y_i) \oplus L_2(x_{i+1})]\} \label{eq:mlca-att-final}
\end{align}

\section{Experimental setup}
\label{sec:expt-setup}

\textbf{Dataset:} Experiments are conducted on the LA partition of ASVspoof 2019 corpus \cite{todisco2019asvspoof}, which provides a standard dataset for anti-spoofing. The LA partition consists of bonafide audios and spoofed audios generated by different TTS and VC algorithms. The training and development subsets share the same attack algorithms, while the evaluation subset utilizes 11 unseen attacks (A07-A15, A17 and A18) and two attacks (A16 and A19) from the training set but trained with different data. The detailed information is shown in Table~\ref{tab:asvspoof2019-corpus}. Systems are evaluated by the tandem detection cost function (t-DCF) \cite{todisco2019asvspoof} and equal error rate (EER) \cite{todisco2019asvspoof}. The log-probability of the bonafide class is adopted as the score for t-DCF and EER computation.

\textbf{Model configuration \& Feature extraction:} As an extension work of \cite{li2021replay}, this work adopts the best single system on the LA attacks in \cite{li2021replay}, i.e. Res2Net50 with squeeze-and-excitation (SE) block, as the baseline. The proposed three gating mechanisms are separately integrated into the baseline model for performance comparison. This work also leverages ResNet50 with SE block for comparison. As indicated in Fig.~\ref{fig:network_structure} (a) and (b), all ResNet, Res2Net and CG-Res2Net models in this work integrate the SE block without explicit denotation. The hyper-parameter $s$ in the Res2Net block and $r$ in the MLCG mechanism are both experimentally set as 4. For acoustic features, this work adopts the constant-Q transform (CQT), which achieves the best results when incorporating with SERes2Net50 \cite{li2021replay}. The CQT is extracted with 16ms step size, Hanning window, 9 octaves with 48 bins per octave. All samples are truncated along the time axis to reserve exactly 400 frames. The samples with less than 400 frames would be extended by repeating their contents.

\begin{table}[t]
\caption{Summary of the ASVspoof 2019 logical access corpus}
\label{tab:asvspoof2019-corpus}
\centering
\begin{tabular}{c|c|c|c}
         \hline
         \hline
         & \#Bonafide & \#Spoofed & Attack algorithms\\
         \hline
        Train & 2,580 & 22,800    & A01-A06 \\
         Dev. & 2,548 & 22,296    & A01-A06 \\
        Eval. & 7,355 & 63,882    & A07-A19 \\
         \hline
         \hline
\end{tabular}
% \vspace{-0.3cm}
\end{table}

\textbf{Training strategy:} The training strategy is identical with \cite{li2021replay}, where binary cross entropy is used to train all models. Adam \cite{kingma2014adam} is adopted as the optimizer with $\beta_1=0.9$, $\beta_2=0.98$ and initial learning rate being $3 \times 10^{-4}$. All models are trained for 20 epochs, and the model with lowest EER on development set is chosen to be evaluated.

\begin{figure*}[t]
    \centering
    \includegraphics[width=\textwidth]{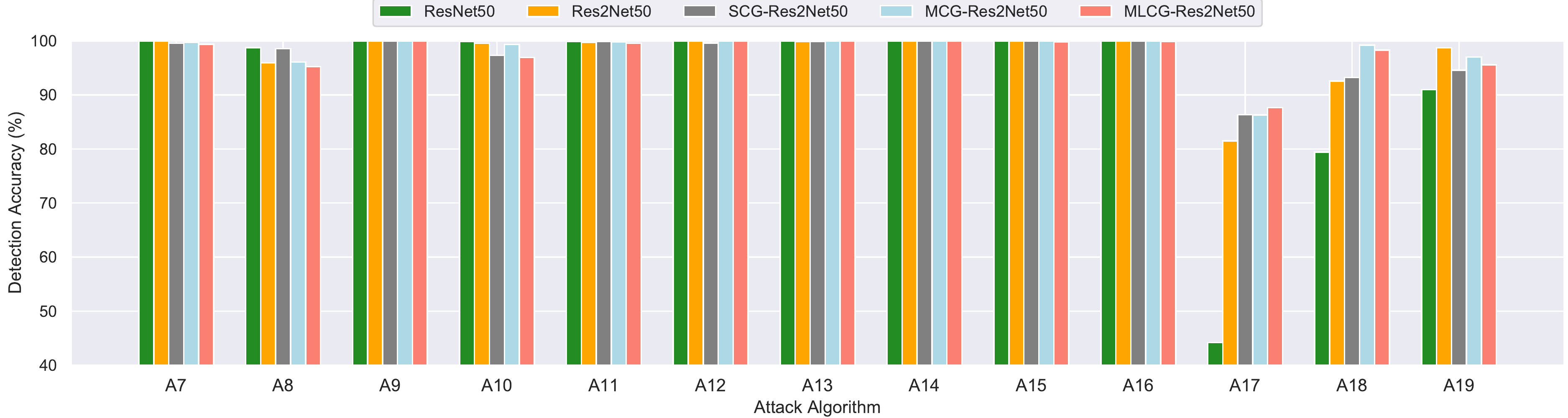}
    \caption{The detection accuracy on each attack, given different network architectures with EER operating points. A16 and A19 are two attacks from the training set but trained with different training data. A17 is the most difficult unseen attack \cite{todisco2019asvspoof}.}
    \label{fig:trr-each-attack}
    % \vspace{-0.3cm}
\end{figure*}

\section{Results}
\label{sec:expt-rst}

\subsection{Effectiveness of channel-wise gated Res2Net}

This section evaluates the effectiveness of proposed CG-Res2Net models for enhancing the generalizability to unseen attacks. Table~\ref{tab:architecture-rst} shows the EER and t-DCF performance of different systems. Notice that the robustness of a spoofing countermeasure depends on its effectiveness on detecting unseen attacks in the evaluation set. Consistent with \cite{li2021replay}, we observe that Res2Net50 performs much better than ResNet50 due to its efficient residual-like connection between feature groups. Compared with Res2Net50, all three CG-Res2Net models demonstrate a superior detection performance on unseen attacks in the evaluation set. Specifically, SCG-Res2Net50 performs slightly better than Res2Net50, while MCG-Res2Net50 and MLCG-Res2Net50 both show a significant improvements over Res2Net50. MCG-Res2Net50 achieves the most promising performance and outperforms Res2Net50 by a relative EER reduction of 28.8\% and a relative t-DCF reduction of 29.7\%. Such results verify the effectiveness of the proposed gating mechanisms, and the necessity of considering the reference information ($x_{i+1}$ in Fig.~\ref{fig:network_structure}d and Fig.~\ref{fig:network_structure}e) in the gating module. MLCG-Res2Net50 did not outperform MCG-Res2Net50 on the overall attacks in the evaluation set, but it has better generalizability to the most difficult unseen attack (A17) in the evaluation set, as will be discussed in Section~\ref{subsec:detection-performance-on-each-algorithm}. Finally, it is also observed that the model complexity of CG-Res2Net models is comparable to that of Res2Net50 and smaller than that of ResNet50, which verifies the efficiency of the proposed gating mechanisms.

\begin{table}[t]
\caption{The EER (\%) and t-DCF of different network architectures on the ASVspoof 2019 logical access.}
\label{tab:architecture-rst}
\centering
\resizebox{0.46\textwidth}{11mm}{
\begin{tabular}{c|c|c|c|c|c}
         \hline
         \hline
         \multirow{2}{*}{System} & \multirow{2}{*}{\# params} & \multicolumn{2}{|c}{Dev. Set} & \multicolumn{2}{|c}{Eval. Set}  \\
         \cline{3-6}
          & & EER (\%) & t-DCF & EER (\%) & t-DCF \\
          \hline
          ResNet50 & 1.09M & 1.09 & 0.037 & 6.70 & 0.177 \\
          Res2Net50 & 0.92M & \textbf{0.43} & \textbf{0.014} & 2.50 & 0.074 \\
          \hline
          SCG-Res2Net50 & 0.95M & 0.59 & 0.018 & 2.43 & 0.076 \\
          MCG-Res2Net50 & 0.96M & 0.47 & 0.015 & \textbf{1.78} & \textbf{0.052} \\
          MLCG-Res2Net50 & 0.94M & 0.86 & 0.027 & 2.15 & 0.069 \\
         \hline
         \hline
\end{tabular}}
% \vspace{-0.3cm}
\end{table}

\subsection{Detection performance on each unseen attack}
\label{subsec:detection-performance-on-each-algorithm}
As mentioned in Section~\ref{sec:expt-setup}, the LA evaluation set has 11 unseen attacks (A07-A15, A17 and A18) and two attacks (A16 and A19) from the training set but trained with different data. As reported in the official ASVspoof 2019 summary \cite{todisco2019asvspoof}, A17 is the most difficult, such that most submitted systems failed to detect it. To perform a detailed system evaluation on each unseen attack, this section reports the detection accuracy on data of each attack, given the system's operating point in terms of EER, as shown in Fig.~\ref{fig:trr-each-attack}.
It has been observed that A17 and A18 are two difficult unseen attacks. In particular, for the most difficult A17 attack, ResNet50 has a detection accuracy below 50\% and Res2Net50 only achieves an accuracy of 81.48\%. All the three proposed CG-Res2Net50 models outperform Res2Net50 by a large margin. MLCG-Res2Net50 achieves the highest detection accuracy of 87.63\%, which outperforms Res2Net50 by 6.15\% in absolute accuracy. For the A18 attack, ResNet50 and Res2Net50 achieve a detection accuracy of 79.41\% and 92.55\%, respectively. The proposed MCG-Res2Net50 and MLCG-Res2Net50 outperform them with a detection accuracy of 99.21\% and 98.27\%, respectively. These observations verify the effectiveness of MCG-Res2Net50 and MLCG-Res2Net50 on generalization to difficult unseen attacks. For other easily detectable attacks, CG-Res2Net50 models perform comparably well with Res2Net50.

\subsection{Comparison with the state-of-the-art single systems}
This section compares the proposed CG-Res2Net models with some reported single, SOTA systems evaluated on the ASVspoof 2019 LA partition, including systems that are submitted to the ASVspoof 2019 competition and those reported in works afterwards (according to our best knowledge). The EER and t-DCF performance are shown in Table~\ref{tab:system-comparison}. The systems are denoted by a name that encodes the input features, system architecture and loss criteria. 

We observe that existing efforts dedicated into acoustic features and data augmentation \cite{das2021data,wu2020light,alzantot2019deep}, system architecture \cite{tak2021end,li2021replay,gomez2020kernel} and loss criteria \cite{gomez2020kernel} have achieved very promising performance. As an extension of \cite{li2021replay}, the proposed CG-Res2Net models outperform other SOTA systems, depicting the effectiveness of the gating mechanisms within the Res2Net block. Moreover, the proposed CG-Res2Net models can be utilized as a backbone network, to be integrated with other effective strategies, e.g. loss criteria, for stronger generalization to unseen attacks.

\begin{table}[t]
\caption{Performance comparison of CG-Res2Net models to some known state-of-the-art single systems on the ASVspoof 2019 LA evaluation set.}
\label{tab:system-comparison}
\centering
\resizebox{0.46\textwidth}{30mm}{
\begin{tabular}{c|c|c}
         \hline
         \hline
         System & EER (\%) & t-DCF \\
         \hline
        Spec+ResNet+CE\cite{lai2019assert} & 11.75 & 0.216 \\
         Spec+ResNet+CE\cite{alzantot2019deep} & 9.68 & 0.274 \\
         MFCC+ResNet+CE\cite{alzantot2019deep} & 9.33 & 0.204 \\
         CQCC+ResNet+CE\cite{alzantot2019deep} & 7.69 & 0.217 \\
          LFCC+LCNN+A-softmax\cite{lavrentyeva2019stc} & 5.06 & 0.100 \\
           FFT+LCNN+A-softmax\cite{lavrentyeva2019stc} & 4.53 & 0.103 \\
        RawAudio+RawNet2+CE\cite{tak2021end} & 4.66 & 0.129 \\
        FG-CQT+LCNN+CE\cite{wu2020light} & 4.07 & 0.102 \\
        
        Spec+LCGRNN+GKDE-Softmax\cite{gomez2020kernel} & 3.77 & 0.084 \\
        Spec+LCGRNN+GKDE-Triplet\cite{gomez2020kernel} & 3.03 & 0.078 \\
        DASC-CQT+LCNN+CE\cite{das2021data} & 3.13 & 0.094 \\
        CQT+SERes2Net50+CE\cite{li2021replay} & 2.50 & 0.074 \\
         \hline
         \textbf{Ours: CQT+SCG-Res2Net50+CE} & 2.43 & 0.076 \\
         \textbf{Ours: CQT+MCG-Res2Net50+CE} & \textbf{1.78} & \textbf{0.052} \\
         \textbf{Ours: CQT+MLCG-Res2Net50+CE} & 2.15 & 0.069 \\
         \hline
         \hline
\end{tabular}}
% \vspace{-0.3cm}
\end{table}

\section{Conclusion}
\label{sec:conclusion}
This work proposes a novel network architecture, i.e. CG-Res2Net, to enhance the model's generalization to unseen attacks. It modifies the Res2Net block to enable a channel-wise gating mechanism in the residual-like connection between feature groups. Such a gating mechanism dynamically selects channel-wise features based on the input, to suppress the less relevant channels and enhance the detection generalizability. Three gating mechanisms are proposed and verified to be effective in enhancing generalization. In terms of overall performance on the LA evaluation set, MCG-Res2Net achieves the best performance and outperforms the Res2Net by a relative EER reduction of 28.8\% and a relative t-DCF reduction of 29.7\%. On the most difficult unseen attack (A17), MLCG-Res2Net achieves the best performance, which outperforms Res2Net by 6.15\% absolute detection accuracy. The proposed CG-Res2Net models outperform other single, SOTA systems on the ASVspoof 2019 LA evaluation, depicting the effectiveness of our method. Given the superior generalizability of CG-Res2Net, future work will investigate its applicability to other speech applications.

\section{Acknowledgement}
We sincerely thank Mr. Zhiyuan Peng from the Chinese University of Hong Kong for some meaningful discussions. This work is supported by HKSAR Government’s Research Grants Council General Research Fund (Project No. 14208718).

\bibliographystyle{IEEEtran}

\bibliography{refs}

% \begin{thebibliography}{9}
% \bibitem[1]{Davis80-COP}
%   S.\ B.\ Davis and P.\ Mermelstein,
%   ``Comparison of parametric representation for monosyllabic word recognition in continuously spoken sentences,''
%   \textit{IEEE Transactions on Acoustics, Speech and Signal Processing}, vol.~28, no.~4, pp.~357--366, 1980.
% \bibitem[2]{Rabiner89-ATO}
%   L.\ R.\ Rabiner,
%   ``A tutorial on hidden Markov models and selected applications in speech recognition,''
%   \textit{Proceedings of the IEEE}, vol.~77, no.~2, pp.~257-286, 1989.
% \bibitem[3]{Hastie09-TEO}
%   T.\ Hastie, R.\ Tibshirani, and J.\ Friedman,
%   \textit{The Elements of Statistical Learning -- Data Mining, Inference, and Prediction}.
%   New York: Springer, 2009.
% \bibitem[4]{YourName17-XXX}
%   F.\ Lastname1, F.\ Lastname2, and F.\ Lastname3,
%   ``Title of your INTERSPEECH 2021 publication,''
%   in \textit{Interspeech 2021 -- 20\textsuperscript{th} Annual Conference of the International Speech Communication Association, September 15-19, Graz, Austria, Proceedings, Proceedings}, 2020, pp.~100--104.
% \end{thebibliography}

\end{document}